\documentstyle[epsf,epsfig,aps]{revtex}
\begin{document}
\preprint{\vbox{\hbox{DOE/ER/40762-177}\hbox{UMD PP\#99-098}}}

\title {Signatures of Disoriented Chiral Condensates from Charged Pions}
\author{Thomas D.~Cohen and Chi-Keung Chow}
\address{Department of Physics, University of~Maryland, College~Park, 
MD~20742-4111}
\date{\today}
\maketitle
\begin{abstract}
We show that the variance in the number of charged pions (in a suitable range 
of momentum space) provides a signature for the observation  of a disoriented
chiral condensate (D$\chi$C).  
The signal should be observable even if multiple domains of D$\chi$C form 
provided the average number of pions per domain is significantly large than 
unity.  
The variance of the number charged pions alone provides a signal which can be 
used even if the number of neutral pions cannot be measured in a given 
detector.  
If the neutrals can be measured, however, the fluctuations in the total number 
of pions provides a signature
which distinguishes disoriented chiral condensates from other hypothetical 
sources of coherent states of pions.
\end{abstract}
\section{introduction}

During the past several years, there has been considerable excitement about the
possibility of the formation of disoriented chiral condensates (D$\chi$C) in 
heavy ion collisions 
\cite{And,KK,An1,An2,BK,Bj1,Bj2,RW1,KT,RW2,BKT1,BKT2,GGP,GM,CBNJ,CKMP,AHW,R,C,Bj3}.  
The basic scenario is as follows:  In an ultrarelativistic heavy ion collision 
some region thermalizes at a temperature above the chiral restoration 
temperature.  
If the system cools sufficiently rapidly back through the transition 
temperature, the region will remain in a chiral restored phase.
However, this phase is unstable; small fluctuations in any chiral direction
($\sigma,\vec{\pi}$) will grow exponentially.  
This can create regions where the pion field has macroscopic occupation.
It should be stressed that this scenario is not derivable directly from the
underlying theory of QCD and contains a number of untested dynamical
assumptions, principally that the cooling is rapid.  
Thus the failure of the system to form a D$\chi$C cannot be used to rule out 
that the system has reached the chiral restoration temperature.  
On the other hand,  observation of the formation of a D$\chi$C would be clear 
evidence that the phase transition had been reached.  
  
Unfortunately, since the scenario is not derived directly from a 
well-defined theory, it is difficult to know precisely what constitutes 
observation of a D$\chi$C.
Assuming the system forms a single large domain of D$\chi$C containing a large 
number of pions there should be clear signatures.  
In the first place, one expects an excess in the number of low $p_T$ pions 
produced.
They would be at low $p_T$ since, by hypothesis, the region is large so the 
characteristic momentum is small; the excess would be measured relative 
to a purely statistical thermal distribution.  
Such a signal has the advantage of working even if multiple regions of 
D$\chi$C form provided each region is large enough so the characteristic 
momentum is sufficiently small to provide a discernible signal over the 
thermal background.  
Such a signal is not decisive since one could imagine some other 
collective low energy effects which produce low $p_T$ pions.    

A much stronger signal of a single large domain of D$\chi$C has been proposed. 
Since the pions formed in a D$\chi$C are essentially classical they form a 
coherent state.  
The coherent state has some orientation in isospin space (or more 
precisely the system is a quantum superposition of coherent states with 
different orientations and particular correlations to the isospin of the 
remainder of the system \cite{CBNJ,C}).  
In essence all of the pions in the domain are pointing in the same 
isospin direction.  
Provided the total number of pions in the domain is large, this implies 
that the distribution of the ratio of neutral to total pions in the domain is 
given by \cite{And,KK,An1,An2,Bj1,Bj2,RW1,R} 
\begin{equation}
f(R) \, =  \, \frac{1}{2 \sqrt{R}}
\label{PofR}
\end{equation}
where R is the ratio of the number of $\pi_0$'s in the D$\chi$C divided
 by the total number of pions and $P(R)$ is the probability.
The derivation of $P(R)$ is quite simple and will be discussed below.  
The distribution in Eq.~(\ref{PofR}) is qualitatively distinct from 
a purely statistical distribution in which the emission of charged and neutral 
pions is uncorrelated.  
The distribution from uncorrelated emissions in the infinite particle number 
limit approaches a delta function at $R=1/3$.  
For finite (but large) particle number the statistical distribution is 
narrowly peaked about 1/3 with a variance, $\langle R^2 \rangle - \langle R 
\rangle^2 = \frac{2}{9{\cal N}}$ where $\cal N$ is the total number of pions.  
Since these two distributions are so radically different one should in 
principle have a very clear signal if a single region of D$\chi$C where to 
form in heavy ion reactions and if the pions from the D$\chi$C are 
kinematically separated from other pions in the system.  

The dramatic nature of the preceding signature is based in large measure
on the assumption that a single large domain of D$\chi$C is formed.  
{\it A priori} this seems rather unlikely for the following reason: 
If a large region of the system starts in a hot chirally restored phase and 
then rapidly cools through the phase transition, then there will be a large 
region which is unstable against growth of the pion field.
Presumably, this happens as a ``seed'' fluctuation in a small region which 
rapidly grows.  
It takes a time of at least $L/c$ for information about the formation of the 
domain to propagate a distance $L$.  
However during the time this fluctuation is growing out to $L$, the pion field 
at $L$ has been sitting in an unstable situation.  
The characteristic time it can remain in this unstable configuration is 
$\tau$, the exponential growth time.  If the information about the initial 
seed does not reach $L$ is a time comparable to $\tau$ the region near $L$ 
will likely begin its own exponential growth but in a chiral direction 
uncorrelated from the initial growth.  
Thus, one expects domains of characteristic size $c\tau$ \cite{GGP,GM}.  

The effect of multiple domains on the $R$ distribution is fairly clear: it will
tend to wash out the signal.  
If a large number of domains form and the pions emerging from different 
domains cannot be distinguished kinematically it is clear from the central 
limit theorem that the $R$ distribution will approach a normal distribution.  
This normal distribution may be distinguished from the normal distribution 
arising from uncorrelated emission; the case of multiple domains of D$\chi$C 
will have a substantially larger variance.

Unfortunately, there is an important practical limitation which makes it 
difficult to exploit the $R$ distribution as a signature.   
Even under the most optimistic of scenarios, the total number of pions coming 
from D$\chi$C's will be a small fraction of the total number of pions.  
If one includes all pions produced in the reaction, the signal from the pions 
from the D$\chi$C will presumably be overwhelmed.  
Thus, it is highly desirable to use kinematic consideration to enhance the 
contributions coming from the D$\chi$C.  
In particular, it is sensible to study the $R$ distribution for a sample 
restricted to low $p_T$ pions only.
In any scenario where the D$\chi$C is well defined, {\it i.e.}, the occupation 
number is large is likely to require a moderately large regions of D$\chi$C 
and the characteristic momentum spread in the D$\chi$C will be fixed by the 
inverse size of the region.  
Thus one expects D$\chi$Cs to preferentially produce moderately low $p_T$ 
pions.  
(One also should restrict the pions in the distribution to a moderately 
narrow rapidity window). 

As an experimental matter, it should be relatively straightforward to cut on 
the momentum of the charged pions to select low $p_T$ pions in a given 
rapidity window.  
For neutral pions, however, it is not a simple matter.  
The neutral pions will decay in flight and will ultimately be detected as 
photons.  
If one is simply interested in the overall $R$ distribution, without cuts, 
and if the detected photons come predominately from $\pi^0$ decays then one 
can use the $n_\gamma/2$ as a surrogate for $n_{\pi^0}$.  
Recent experimental searches have exploited this strategy\cite{MM,Bj4,WA}.  
However, in order to study the $R$ distribution in a limited kinematical 
region it is necessary to reconstruct the $\pi^0$ momenta from the observed 
photons in order to make kinematical cuts on the $\pi^0$ momenta.  
Since the number of neutral pions per event is large, the reconstruction of 
neutral pion momenta is likely to be a formidable task.

This raises the following interesting question:  
Can one find a signature for the presence of regions of D$\chi$C of 
essentially the same quality as the $R$ distribution but which does not 
require the measurement of neutral pions?
In this article, we will show that the distribution of the number of
charged pions (in a kinematically limited region) contains essentially the 
same information about D$\chi$C formation as the the $R$ distribution.  
This should greatly aid in searches for D$\chi$C formation.

We also discuss additional information that can be inferred if $\pi^0$'s can
be reconstructed.  
In particular, we show that the distribution of the total number of pions (in 
a limited kinematic region) provides a means to distinguish D$\chi$C formation 
from other hypothetical mechanisms for the production of a pion coherent 
state.

\section{A Simplified Model}

We begin by studying an overly simple model and in subsequent sections we will
generalize our results to more realistic scenarios.  
In this simplified situation we assume that in every collision a single large 
domain of D$\chi$C is formed with a large particle number.  
Moreover, we will assume that the field strength and spatial distribution of 
this domain do not vary event by event, and that the pions produced in the 
D$\chi$C are kinematically completely distinguishable from all other pions in 
the system (including those pions produced from ``$\sigma$'s'' --- {\it i.e.}, 
fluctuations in the $\langle \overline{q} q \rangle$ directions).  
Finally, we will assume that both isospin violating effects and explicit 
chiral symmetry breaking are negligible.  

By hypothesis, the region of D$\chi$C contains many particles and is
essentially classical in nature. 
To simplify discussion we will adopt the usual convention of 
describing the physics in terms of the degrees of freedom in a linear sigma 
model with O(4) symmetry, {\it i.e.}, $\sigma$ and $\vec{\pi}$ 
rather than directly in terms of the QCD degrees of freedom. 
We wish to stress, however, that we are not relying on the detailed dynamics 
of any particular variant of the $\sigma$ model.

Being a coherent state, the D$\chi$C can be written in the following form: 
\begin{equation}
|\hbox{D$\chi$C}(\psi,\theta,\phi)\rangle = \exp
\left(\int {d^3k \over (2\pi)^3}\; f(k) \vec a^\dag \cdot \vec n\right) 
|{\rm vac}\rangle, 
\end{equation}
where $\vec a^\dag = (\pi_x^\dag, \pi_y^\dag, \pi_0^\dag, \sigma^\dag)$ is 
the vector of creation operators in the chiral space. 
The unit vector $\vec n=(\sin\psi \sin\theta \cos\phi, 
\sin\psi \sin\theta \sin\phi, \sin\psi \cos\theta, \cos\psi)$ denotes the 
orientation of the D$\chi$C in the chiral space, $f(\vec k)$ is the 
distribution of the D$\chi$C in momentum space, and $|{\rm vac}\rangle$ is 
the vacuum.  

The number operators for neutral and charged pions are 
\begin{equation}
n_0 = \pi_0^\dag \pi_0, \qquad n_\pm = \pi_x^\dag \pi_x + \pi_y^\dag \pi_y, 
\end{equation}
and one can easily find their expectation values , {\it i.e.}, $\langle 
n_{0,\pm} \rangle(\psi,\theta,\phi) = \langle \hbox{D$\chi$C}
(\psi,\theta,\phi)| n_{0,\pm} |\hbox{D$\chi$C}(\psi,\theta,\phi)\rangle$. 
It turns out that $\langle n_{0,\pm}\rangle$ can be factorized into the 
following form.  
\begin{equation}
\langle n_{0,\pm} \rangle(\psi,\theta,\phi) = \langle n \rangle \, 
g_{0,\pm}(\psi,\theta,\phi), 
\label{fac}
\end{equation}
where $\omega^2 = \vec k^2 + m_\pi^2$, and 
\begin{equation}
\langle n \rangle = \langle \vec a^\dag \cdot \vec a \rangle = 
\int {d^3k\over (2\pi)^3}\;{f^2(\vec k)\over\omega}.    
\end{equation}
The expectation value $\langle n \rangle$ measures the total number of 
$\pi$'s produced by the D$\chi$C if fully oriented in a pionic direction.  
The geometrical factors $g_0(\psi,\theta,\phi)$ and $g_\pm(\psi,\theta,\phi)$ 
will be called the neutral and charged proportions respectively, and they take 
the following forms.  
\begin{equation}
g_0(\psi,\theta,\phi) = \sin^2 \psi \cos^2 \theta, \qquad 
g_\pm(\psi,\theta,\phi) = \sin^2 \psi \sin^2 \theta.
\label{defg}
\end{equation}
Note that $\langle n \rangle$ does not depend on the 
orientation angles $(\psi, \theta, \phi)$ while $g_{0,\pm}$ do not depend on 
the dynamical variables $f(k)$ and $\omega$.  

One can also evaluate the higher moments of $n_{0,\pm}$, which gives 
\begin{equation}
\langle n_{0,\pm}^2 \rangle(\psi,\theta,\phi) = \langle n^2\rangle \,
g^2_{0,\pm}(\psi,\theta,\phi),
\end{equation}
with 
\begin{equation}
\langle n^2\rangle = \langle n \rangle^2 + \langle n \rangle. 
\end{equation}
Let's define the {\it deviance\/} $\delta[X]$ of a distribution of variable 
$X$ such that 
\begin{equation}
\langle X^2 \rangle = (1+\delta[X]) \langle X \rangle^2, \qquad \hbox{or} 
\qquad (\Delta X)^2 \equiv \langle X^2 \rangle - \langle X \rangle^2 
= \delta[X] \langle X \rangle^2.   
\end{equation}
In this case, 
\begin{equation}
\delta[n] = 1/\langle n \rangle \to 0 \quad \hbox{when $\langle n \rangle \to 
\infty$}.  
\end{equation}  
Actually it is straightforward to show that $n$ is described by a Poisson 
distribution, which always has a small deviance as the variance is 
proportional to the mean.  

\bigskip

The above analysis shows that, {\it for each set of orientation angles 
$(\psi, \theta, \phi)$,} the distributions of $n_{0,\pm}$ are normal.  
However, since the orientation is randomly generated in the process of 
spontaneous symmetry breaking, one cannot predict $(\psi, \theta, \phi)$.  
On the other hand, since we are neglecting explicit chiral symmetry breaking, 
the system is equally probable to point in any direction in chiral space.  
Moreover, since by hypothesis we are in a semiclassical situation (large 
$\langle n \rangle$), it is legitimate to work with probabilities rather 
than amplitudes.  
Using the technology of Ref.~\cite{R}, the probability 
distribution in the angular variables is given by the unit measure:  
\begin{equation}
d^3P(\psi,\theta,\phi) = \frac{1}{2 \pi^2} \; \sin^2\psi \, \sin\theta \; 
d\psi \, d\theta \, d\phi .  
\label{pa}
\end{equation}
One can use Eq.~(\ref{defg}) to 
reparametrize the probability distribution (\ref{pa}) in terms of the 
neutral and charged proportions $g_{0,\pm}=\langle n_{0,\pm}\rangle/
\langle n \rangle$, {\it i.e.}, the fraction of neutral or charged pions among 
all particles produced by the D$\chi$C.  
\begin{equation}
d^2P(g_0,g_\pm) = {1\over \pi} \, {1\over \sqrt{g_0(1-g_0-g_\pm)}} 
\;dg_0\, dg_\pm.
\label{j}
\end{equation}
From this one can obtain the marginal probability distribution of $g_0$ and 
$g_\pm$ by integrating over the other variable.  
\begin{eqnarray}
dP(g_0) & = f_0(g_0) dg_0 = & {2\over \pi} \, \sqrt{1-g_0\over g_0} dg_0, 
\nonumber\\
dP(g_\pm) & = f_\pm(g_\pm) dg_\pm = & dg_\pm.  
\end{eqnarray}
These distribution functions are plotted in Fig.~1 in solid curves.  
It is obvious the both distributions are far from being normal.  
The function $f_0$ is heavily skewed towards the low end and actually 
diverges as $1/\sqrt{g_0}$ when $g_0\to 0$.   
On the other hand, $f_\pm$ is flat, and $g_\pm$ is equally likely to assume 
any value between 0 and $1$.  
This is drastically different from pion emission from an uncorrelated source
(the dotted curves in Fig.~1 \footnote{
For uncorrelated emissions, the probability distributions are Poisson-Gaussian 
with mean $1/4$ and $1/2$ for neutral and charged pions respectively.  
The variances depend on the number of independently emitted pions; the 
plots correspond to the case of $n=50$.}), 
where both distributions would be normal.  

\bigskip
\begin{figure}
\epsfig{file=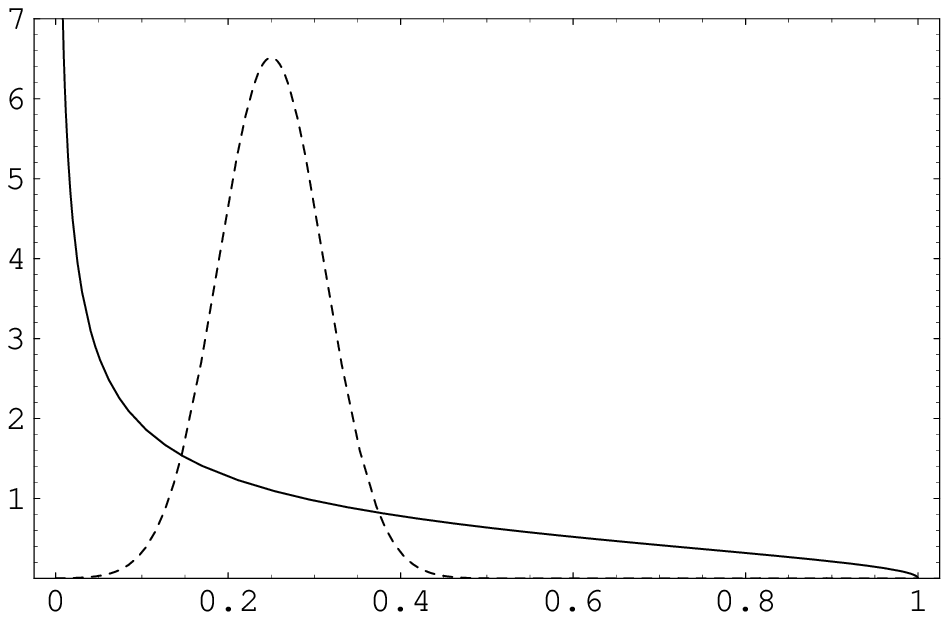, width=3.4in} 
\epsfig{file=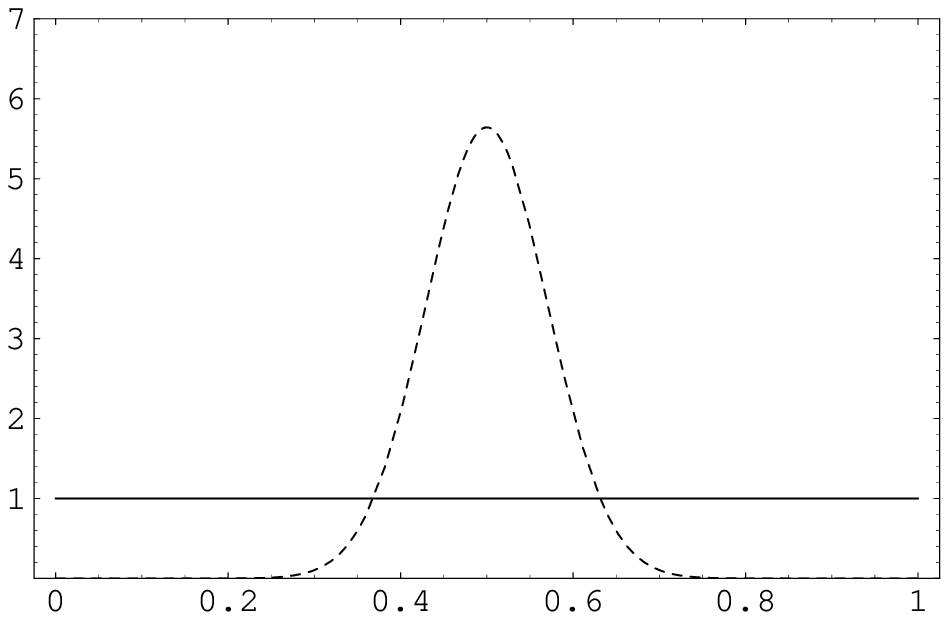, width=3.4in}

\bigskip

\caption{The probability distribution functions $f_{0,\pm} (g_{0,\pm})$.  
The plot on the left is $f_0(g_0)$, and the right is $f_\pm(g_\pm)$.  
The solid curves are for DCC emission, while the dotted curves are for 
independent emission with $n=50$.  }
\end{figure}
\bigskip

More quantitatively, one can calculate the first and second moments of 
$g_0$ and $g_\pm$.  
\begin{eqnarray}
\langle g_0 \rangle &= 1/4, \qquad\qquad \langle g_0^2 \rangle =& 1/8, 
\nonumber\\
\langle g_\pm \rangle &= 1/2, \qquad\qquad \langle g_\pm^2 \rangle =& 1/3,   
\end{eqnarray}
and hence the respective deviance $\delta[g_0]$ and $\delta[g_\pm]$, 
\begin{equation}
\delta[g_0] = 1, \qquad \delta[g_\pm] = 1/3.  
\end{equation} 

What does the distributions of the proportions $g_{0,\pm}$ tell us about the 
distributions of $n_{0,\pm}$?  
It is obvious that when $n$ is fixed, $g_{0,\pm}$ give the pion distribution. 
In reality, of course, $n$ is not fixed; we have shown that it behaves like a 
Poisson distribution if we have a conventional Glauber coherent state 
\cite{G}.  
However, since Poisson distributions are sharply peaked, the dispersion of 
$n$ will simply smear the distribution of $n_{0,\pi}$ slightly, without 
changing the overall shape qualitatively.  
More specifically, note that \footnote{
The two set of brackets in the right-hand side of the equations below have 
different physical origins.  
The expectations of $g_{0,\pm}$ are statistical in nature, while that of 
$n$ is via quantum mechanical smearing of a coherent state.  
Consequently the distributions of $n$ and $g_{0,\pm}$ are assumed to be 
uncorrelated, and we derived the equalities below.}  
\begin{eqnarray}
\langle n_{0,\pm} \rangle &=& \langle n \rangle \, 
\langle g_{0,\pm} \rangle, \nonumber\\
\langle n_{0,\pm}^2 \rangle &=& \langle n^2 \rangle \, 
\langle g_{0,\pm}^2 \rangle = (1+\delta[n])\, (1+\delta[g_{0,\pm}]) \; 
(\langle n \rangle \, \langle g_{0,\pm} \rangle)^2 \equiv (1+\delta[n_{0,\pm}])
\langle n_{0,\pm} \rangle^2.  
\end{eqnarray}
The first equality gives 
\begin{equation}
\langle n_0 \rangle = \langle n \rangle / 4, \qquad 
\langle n_\pm \rangle = \langle n \rangle / 2,  
\end{equation}
which have the simple interpretation that, given the symmetry between the 
four directions $(\pi_x, \pi_y, \pi_0, \sigma)$ in the chiral space, 
a quarter of the particles produced by the D$\chi$C will be $\pi_0$, while 
half of them will be $\pi_x$ or $\pi_y$.  
On the other hand, the second equality gives 
\begin{equation}
\delta[n_{0,\pm}] = \delta[n] + \delta[g_{0,\pm}] + \delta[n]\delta[g_{0,\pm}].
\end{equation}
With $\delta[n] = 1/\langle n\rangle$, $\delta[g_0]=1$ and $\delta[g_\pm]$, 
one has 
\begin{equation}
\delta[n_0] = 1 + 2/\langle n \rangle, \qquad 
\delta[n_\pm] = 1/3 + 4/3\langle n \rangle.  
\label{ds}
\end{equation}
As expected, when $\langle n \rangle$ is large, the deviances 
$\delta[n_{0,\pm}]$ approach $\delta[g_{0,\pm}]$.  
Note that both $\delta[n_{0,\pm}]$ are of order 1, in contrast to an 
uncorrelated emission which would have $\delta = 1/\langle n \rangle$.    
In an ideal world, such enhancements of $\delta$'s would indicate the 
existence of D$\chi$C.  

Lastly, it is also useful to reproduce the aforementioned probability 
distribution of $R = n_0/(n_0+n_\pm) = g_0/(g_0+g_\pm)$ by reparametrizing 
distribution (\ref{j}) in terms of $R$ and $g_t=g_0+g_\pm$ (the subscript 
$t$ stands for total). 
\begin{equation}
d^2 P(g_t,R) = {1\over \pi} \, \sqrt{g_t\over R(1-g_t)}\;dg_t\,dR, 
\end{equation}
with the marginal distributions 
\begin{eqnarray}
dP(g_t)&= f_t(g_t)\,dg_t=&{2\over\pi}\sqrt{g_t\over 1-g_t}\,dg_t,\nonumber\\
dP(R)&= f_R(R)d_R = &{1\over 2\sqrt{R}}dR.  
\end{eqnarray}
Note that while $f_0$ is drastically skewed towards the low end and 
$f_\pm$ is flat, $f_t$ is skewed towards the high end.  
This may sound counter-intuitive, but one must bear in mind that the emissions 
of neutral and charged pions are not independent and there is no contradiction.
Also note that $f_R$ is exactly as predicted in 
Ref.~\cite{And,KK,An1,An2,Bj1,Bj2,RW1,R} (cf.~Eq.~(\ref{PofR})).

\section{More Realistic Scenarios}

We have studied the simple case of pion emission from a single huge (in 
the sense that $\langle n \rangle$ is large) D$\chi$C domain.  
As discussed in the introduction, this scenario is presumably not realistic.  
We will now proceed to study more realistic scenarios with multi-domain 
formation.  
The main point is, even though the probability distribution is smeared out 
because of the lack of alignment (in the chiral space) between the different 
domains, one feature survives, namely the large variance of the distributions.
In particular, we will see that the variances for both neutral and charged 
productions are still much larger than that of an uncorrelated emission.  

Let's consider the case which we have $N$ domains, each with the same 
$\langle n \rangle$.  
We are also assuming both $\langle n \rangle$ and $N$ are much larger 
than unity.  
The total number of pions of each species produced is the sum of pions 
of that particular species produced in each domain, the distribution of 
which has been discussed in the previous section.  
\begin{equation}
\Sigma n_0 = \sum_{i=1}^N n_0^{(i)},\qquad 
\Sigma n_\pm = \sum_{i=1}^N n_\pm^{(i)}.  
\end{equation}
By the central limit theorem, the probability distribution of $\Sigma 
n_{0,\pm}$ will approach normal distributions when $N$ is large.  
However, we will see that the variances of the Gaussian distributions will be 
much larger for pion production from a D$\chi$C than those of uncorrelated 
pion emission. 

Since the pion production in each domain are independent, $n_{0,\pm}^{(i)}$ 
are independent random variables.  
Hence the mean of $\Sigma n_{0,\pm}$ is just the sum of the means of all  
$n_{0,\pm}^{(i)}$, 
\begin{equation}
\langle \Sigma n_{0,\pm} \rangle = \sum_{i=1}^N \langle n^{(i)}_{0,\pm} 
\rangle = N \langle n_{0,\pm} \rangle \equiv {\cal N}_{0,\pm}, 
\label{mean}
\end{equation}
and the variance of the sum $n_{0,\pm}$ is just the sum of the variances of 
each $n_{0,\pm}^{(i)}$.  
\begin{eqnarray}
(\Delta \Sigma n_{0,\pm})^2 &=& \sum_{i=1}^N (\Delta n^{(i)}_{0,\pm})^2 = 
N (\Delta n_{0,\pm})^2 \nonumber\\ &=& N \langle n_{0,\pm} \rangle^2 
\delta [n_{0,\pm}] = (\delta[n_{0,\pm}] / N) {\cal N}_{0,\pm}^2.    
\label{var}
\end{eqnarray}
In other words, 
\begin{equation}
\delta[\Sigma n_{0,\pm}] = {(\Delta \Sigma n_{0,\pm})^2 \over 
\langle \Sigma n_{0,\pm} \rangle^2} = {\delta[n_{0,\pm}] \over N} = 
{\delta[n_{0,\pm}] \langle n_{0,\pm} \rangle \over {\cal N}_{0,\pm}}.  
\label{wide}
\end{equation}
In comparison with uncorrelated pion production, with $\delta = 
1/{\cal N}_{0,\pm}$ we see that for multi-domain D$\chi$C the deviances are 
enhanced by a factor of $\epsilon_{0,\pm} = \delta[n_{0,\pm}] 
\langle n_{0,\pm} \rangle$.  
\begin{equation}
\epsilon_0 = (\langle n \rangle + 2)/4, \qquad 
\epsilon_\pm = (\langle n \rangle + 4)/6.  
\label{e1}
\end{equation}
{\it A priori\/} $\langle n \rangle$ can take any value, but 
$\epsilon_{0,\pm}$ are larger than unity for any value of $\langle n \rangle 
> 2$.  
Even for a very modest $\langle n \rangle =8 $, $\epsilon_0 = 2.5$ and 
$\epsilon_\pm = 2$, leading to substantial widening of the corresponding 
distributions, an observable signature of D$\chi$C formation.  
For larger values of $\langle n \rangle$, the broadening will be even more 
pronounced.  

\bigskip

While the above scenario describes D$\chi$C with multi-domains, a probable 
feature of D$\chi$C formation in the real world (if it happens at all), it is 
still unrealistic in assuming all the domains are of equal strength, 
{\it i.e.}, with the same $\langle n \rangle$.  
Instead one expects $\langle n \rangle$ of different domains to fall under 
a certain probability distribution, which depends on the details of the 
model.  
Naturally, one questions if the signatures discussed above still survive 
under such circumstances.  

Let's consider the case with $N$ domains, with different $\langle n^{(i)} 
\rangle \gg 1$.  
Equation (\ref{mean}) becomes 
\begin{equation}
\langle \Sigma n_{0,\pm}\rangle = \sum_{i=1}^N \langle n^{(i)}_{0,\pm}\rangle
= N \overline {\langle n_{0,\pm} \rangle} \equiv {\cal N}_{0,\pm}, 
\end{equation}
where $\overline {\langle n_{0,\pm} \rangle}$ is the average of  
$\langle n^{(i)}_{0,\pm}\rangle$.  
Equation (\ref{var}) becomes   
\begin{equation}
(\Delta \Sigma n_{0,\pm})^2 = \sum_{i=1}^N (\Delta n^{(i)}_{0,\pm})^2 = 
\sum_{i=1}^N \langle n^{(i)}_{0,\pm}\rangle^2 \delta[n^{(i)}_{0,\pm}].  
\end{equation}
By the inequalities $\delta[n^{(i)}_{0,\pm}] > \delta[g_{0,\pm}]$ 
(cf.~Eq.(\ref{ds})) and $\sum_1^N \langle n \rangle^2 \leq 
(\sum_1^N \langle n \rangle)^2/N$ (mean of squares is larger than square of 
mean), we have 
\begin{equation}
(\Delta \Sigma n_{0,\pm})^2 > \left( \sum_{i=1}^N \langle n^{(i)}_{0,\pm}
\rangle^2 \right) \delta[g_{0,\pm}] \geq \left( \sum_{i=1}^N \langle 
n^{(i)}_{0,\pm}\rangle \right)^2 \delta[g_{0,\pm}] / N = 
(\delta[g_{0,\pm}]/N) {\cal N}_{0,\pm}^2.  
\end{equation}
In other words, 
\begin{equation}
\delta[\Sigma n_{0,\pm}] = {(\Delta \Sigma n_{0,\pm})^2 \over 
\langle \Sigma n_{0,\pm}\rangle^2} \geq {\delta[g_{0,\pm}] \over N} 
= {\delta[g_{0,\pm}] \overline {\langle n_{0,\pm} \rangle} \over 
{\cal N}_{0,\pm}}.  
\end{equation}
(Compare Eq.~(\ref{wide}).)  
Again, the deviances are much larger than that of uncorrelated emission with 
$\delta = 1/{\cal N}_{0,\pm}$ when $\overline {\langle n \rangle} 
\gg 1$ by the following enhancement factors. 
\begin{equation} 
\epsilon_0 \geq \overline {\langle n \rangle}/4, \qquad 
\epsilon_\pm \geq \overline {\langle n \rangle}/6.  
\label{e2}
\end{equation}
So we see that, even with domains of unequal strengths, the number of 
neutral or charged pions produced by a D$\chi$C will still have a much wider 
distribution than that from independent, uncorrelated emission.  

Lastly, one may also ask if the distribution of $\Sigma n_{0,\pm}$ will 
approach normal distributions when $N$ is large in the case of domains 
with unequal strengths.  
In this case $n^{(i)}$'s do not all fall under the same probability 
distribution and the most simple form of the central limit theorem does 
not apply.  
On the other hand, there are generalized forms of the central limit theorem,   
which state that as long as the probability distributions are sufficiently 
``well behaved'', the sum of $N$ random variables will still fall under a 
normal distribution when $N \to \infty$.  
It is actually possible to argue that the distribution of $\Sigma n_\pm$ 
does approach a normal distribution by the Lindeberg generalization of 
the central limit theorem.  
(See, for example, Sec.~6.E of Ref.~\cite{F}.)  
Whether the same conclusion holds for $\Sigma n_0$ is still an open question.  

\section{Discussion}

Let us recapitulate what we have shown in the previous sections.  
We have shown that one can calculate the probability distribution of the 
number of neutral or charged pions produced as a result of D$\chi$C formation. 
The resultant deviances $\delta$'s are not of the order $1/\cal N$ as in an 
uncorrelated emission, but are instead enhanced by factors $\epsilon$'s 
which are of order $\langle n \rangle$.  
Seeing such enhancements of deviances would be signatures of coherent pion 
productions.  

One can understand the origin of such enhancements of statistical 
fluctuations of the number of neutral or charged pions by considering the 
following analogy.  
Consider two groups of gamblers playing roulette in a casino: $N$ lawyers at 
\$100 tables, and $100 N$ physicists at the \$1 tables, where the odds are 
the same.  
If each lawyer and physicist is given $n$ chips, of \$100 and \$1,  
respectively (so that the total amount given to the lawyers, ${\cal N} = N 
\times 100n$, is the same as that given to the physicists, ${\cal N} = 100N 
\times n$), and is required to bet all of them, the average loss will be the 
same for both groups as long as they are following the same betting 
strategies.  
However, it is easy to see that the statistical fluctuation of the loss 
of the lawyers would be much larger for that of the physicists.  
In other words, the amount of loss, as well as its standard deviation, 
is ``quantized'' in units of the value of the bets.  
The larger the bet, the larger the fluctuation. 
On the other hand, one can also turn the argument around; a discerning 
external observer can deduce, with the knowledge of the gambling strategies, 
the size of the bets of the lawyers from the statistical fluctuations of the 
lawyers' losses, and do likewise for the physicists as well.  

Just as the chips in a single bet of a lawyer share the same fate (either win 
or lose), all the pions in a single D$\chi$C domain share the same orientation 
in the chiral space. 
As a result, the fluctuation of the number of pions in each direction in 
the chiral space is enhanced by a factor proportonal to $\langle n \rangle$, 
the number of pions in each D$\chi$C domain.  
And by reverse argument, one can deduce whether coherent pion emission is 
taking place by measuring the fluctuation of the number of emitted pions.  

Our analysis has assumed that all the pions originate from coherent emissions, 
and each of these coherent states has $\langle n^{(i)} \rangle \gg 1$.  
In the real world, there is contamination from independent pion emissions, 
and the total numbers of neutral or charged pions are the sums of these two 
contributions.  
\begin{equation}
n_{0,\pm} = n_{0,\pm}^{(\rm c)} + n_{0,\pm}^{(\rm inc)}, 
\end{equation}
where ``c'' and ``inc'' stands for ``coherent'' and ``incoherent'', 
respectively.  
The variances of the numbers of coherently produced neutral or charged pions 
are enhanced while those of incoherent production are not.  
\begin{equation}
(\Delta n_{0,\pm}^{(\rm c)})^2 = \epsilon_{0,\pm} \langle n_{0,\pm}^{(\rm c)} 
\rangle , \qquad 
(\Delta n_{0,\pm}^{(\rm inc)})^2 = \langle n_{0,\pm}^{(\rm inc)} \rangle,  
\end{equation}
where $\epsilon_{0,\pm}$ are of order $\langle n \rangle \gg 1$ 
(cf.~Eq.~(\ref{e1}), (\ref{e2})).  
Then one can calculate the variance of the sum of the two contributions.  
\begin{equation}
(\Delta n)^2=(\Delta n_{0,\pm}^{(\rm c)})^2+(\Delta n_{0,\pm}^{(\rm inc)})^2
= \tilde\epsilon_{0,\pm} \langle n \rangle, \qquad 
\tilde\epsilon_{0,\pm} = \chi \epsilon + (1-\chi) = 1+ \chi(\epsilon - 1),   
\end{equation}
with 
\begin{equation}
\chi = \langle n_{0,\pm}^{(\rm c)} \rangle / \langle n_{0,\pm} \rangle, \qquad 
1- \chi = \langle n_{0,\pm}^{(\rm inc)} \rangle / \langle n_{0,\pm} \rangle.
\end{equation}
The parameter $\chi$ measures the fraction of pions which are coherently 
produced: $\chi=1$ when all the pions are from D$\chi$C, while $\chi=0$ when 
all of them are independently emitted.  
Obviously, the more incoherent pions in the sample, the smaller is the 
enhancement factor $\tilde\epsilon_{0,\pm}$.   

While these incoherently emitted pions dilute our signatures for D$\chi$C, 
they have different momentum spectra from those from D$\chi$C.  
D$\chi$C pions, being produced from coherent state, carry low $p_T$.  
The typical $p_T$ is of the order of $1/L$, where $L$ is the size of the 
domain from which the pion originates.  
In contrast incoherently emitted pions can carry high $p_T$.  
Therefore applying a low $p_T$ cut can minimize the noise from incoherent pion 
emissions.  

It is also advantageous to measure the rapidity of the pions and count 
their numbers in narrow rapidity windows.  
Bear in mind that the rapidities of the pions are, up to small dispersions, 
equal to that of the original domain.  
As we have mentioned, it is probable that many domains are formed in a 
single collision, and all these domains may have different rapidities. 
For example, the domains at the surface of the ``fire ball'' are moving 
with high speed relative to the domains at the center.  
By binning the pions according to their rapidities, one can partially 
separate the pions from different domains, and the signals are enhanced as 
a result.  

\bigskip

In summary, we suggest the following procedure in looking for signatures 
of D$\chi$Cs.  

$\bullet$ Count the number of neutral or charged pions {\it event by event\/} 
from heavy ion collision experiments and measure their individual transverse 
momenta and rapidities.  

$\bullet$ Apply a low $p_T$ cut to suppress the noise due to uncorrelated 
pion emission. 

$\bullet$ Bin the events in different rapidity windows.  

$\bullet$ In each rapidity window, plot the number of events {\it vs.~}the 
number of neutral or charged pions in histograms.  

$\bullet$ Evaluate the mean, $\langle n_{0,\pm} \rangle$, and the variance, 
$(\Delta n_{0,\pm})^2$, in each rapidity window.  

$\bullet$ If we find $(\Delta n_{0,\pm})^2$ is substantially larger than 
$\langle n_{0,\pm} \rangle$, then we are seeing signatures from D$\chi$Cs.  

The above procedure allows us to search for signatures from D$\chi$Cs by 
counting only the charged pions.  
This is important as, with our present technology, it is difficult to count 
the number of $\pi_0$'s in a momentum bin, which would mean reconstructing 
all the pions from photons --- a formidable task.  
On the other hand, with great experimental effort, it may be possible to 
count the neutral pions as well in the future.  
In that case, we will be able to distinguish D$\chi$C formation from other 
mechanisms of coherent pion productions.  
For example, one can count $n_t$, the number of pions (both neutral and 
charged) in each rapidity window.  
For D$\chi$C formation, or any other mechanisms of coherent pion productions 
where the field is aligned with a random direction in the four-dimensional 
chiral space $(\pi_x,\pi_y,\pi_0,\sigma)$, the fluctuation of $n_t$ is 
large.  
On the other hand, for mechanisms of coherent pion productions where the 
field is aligned with a random direction in the three-dimensional isospace 
$(\pi_x,\pi_y,\pi_0)$ but without involving the $\sigma$ direction, it is 
straightforward to show that the fluctuation of $n_t$ is small.  

In summary, we have constructed signatures for D$\chi$C formation in heavy 
ion collisions which do not require counting the number of neutral pions.  
Instead, we suggest counting the number of charged pions produced, and 
a large fluctuation would be a signal of D$\chi$C formation.  
We believe these new signatures will be useful in searches for D$\chi$C at 
RHIC and LHC.  

\bigskip

Support of this research by the U.S.~Department of Energy under grant 
DE-FG02-93ER-40762 is gratefully acknowledged.

\end{document}